\documentclass[12pt,preprint]{aastex}
\newcommand{\cs}[3]{{{#3} \brace {#1 #2}}}
\newcommand{\edf}{\ {\mathop{=}\limits^{\rm def}}\ }
\lefthead{Wanas and Morcos } \righthead{Theorems on Null-Paths and
red-Shift}

\begin{document}

\title{Theorems on Null-Paths and Red-Shift\\}
\author{M. I. Wanas\altaffilmark{1} and A.B.Morcos\altaffilmark{2}}
\affil{1-Astronomy Department, Faculty of Science, Cairo
University, Giza, Egypt.} \affil{2-Department  of Astronomy ,
National Research Institute of Astronomy and Geophysics,\\ Helwan,
Cairo, Egypt.} \altaffiltext{1}{e.mail:~wanas@frcu.eun.eg}
\altaffiltext{2}{e.mails:~morcos@frcu.eun.eg \\
$~~~~~~~~~~~~~~~~~$fmorcos@nriag.sci.eg}
\begin{abstract}
In the present work, we prove the validity of two theorems on
null-paths in a version of absolute parallelism geometry. A version
of these theorems has been originally established and proved by
Kermak, McCrea and Whittaker (KMW) in the context of Riemannian
geometry. The importance of such theorems is their use, in
applications, to derive a general formula for the red-shift of
spectral lines coming from distant objects. The formula derived in
the present work, can be applied for both cosmological and
astrophysical red-shifts. It takes into account the shifts resulting
from gravitation, different motions of the source of photons, spin
of the moving particle (photons) and the direction of the line of
sight. It is shown that this formula cannot be derived in the
context of Riemannian geometry, but it can be reduced to a formula
given by KMW under certain conditions.
\end{abstract}

\keywords{ Null-Geodesics, Riemannian Geometry , Absolute
Parallelism Geometry, Null-Paths, Red-Shift }

\section{Introduction}
 In the context of the general theory of relativity
          (GR), the red-shift in the spectra of distant objects is
          a metric phenomena. In other words, knowing the metric
          of space-time, one can calculate the red-shift whether
          it is astrophysical or cosmological.This depends on the
          fact that the trajectory of the photons, in a
          background gravitational field, is assumed to be
          null-geodesic, while the metric of the space-time is the first
          integral of this null-geodesic. An alternative, and
          more general, method for calculating the red-shift is by
          using the results of theorems on null-geodesics, established in Riemannian
          geometry, by  Kermack,McCrea, and
          Whittaker (KMW) in 1933. It is worth of mention that the
          applications of the two methods, in the context of GR,
          give identical results. In both methods, it is assumed
          that the trajectory of a photon in a gravitational field
          is a null-geodesic. This implies the neglect of the
          effect of spin of the photon (spin independent
          trajectory), on its trajectory,  if any. In calculating
          the red-shift, the first method is more easier and direct than
          the second; and since the two methods are equivalent in the context
          of GR, authors usually use the first method neglecting the
          second one.

               Recently, some evidences indicating probable
          dependence of trajectories of spinning particles on
          their spin, were reported. One of these evidences, on the laboratory scale, is the discrepancy
          between  theoretical calculations and the
          results of the COW-experiment(Overhauser and Colella (1974),Colella et al. (1980)
          and Werner et al.(1988). Another
          evidence, on the galactic scale, concerning the
          arrival times of photons, neutrinos (and gravitons!) from
          SN1987A (Schramm and Truran(1990), Weber(1994) and De Rujula(1987)).
          On the other hand, a new path equation, in the parameterized absolute parallelism (PAP) geometry,
          giving the trajectory of a spinning particle in a
          gravitational field ( spin dependent trajectory ), is suggested
         (Wanas(1998)). It is worth of mention that
          the use of this equation has given a satisfactory interpretation of the discrepancy in the COW
           experiment Wanas et al. (2000), and can account for the time delay of spinning particles
           coming from SN1987A(Wanas et al.(2002) ). If we use this equation, to describe the trajectory of photons,
           (spin one massless particle), we encounter a problem
          concerning red-shift calculations. That is, the metric of space-time is not the first integral of
          the new path equation; and the red-shift is no longer a metric
          phenomena.
          Consequently, one can not calculate the red-shift using
          any of the above mentioned methods. One way to solve
          this problem is to develop theorems on the null-path, similar to those of
          KMW-theorems, in order to apply them
          for obtaining the red-shift extracted from spinning
          particles. This is the aim of the present work.\\

          For this aim, we give a brief account on KMW-
          theorems in section 2. In section 3, we review the main
          features of the spin dependent path equation. The validity of the KMW-
          theorems are proved in PAP geometry in section 4. In section 5 we derive a general
          formula for  red-shift, taking into
          account the spin of the particle from which we extract
          the red-shift. The work is discussed in section 6.
\section{KMW- Theorems}

          The following theorems on null-geodesics, were established by
    KMW (Kermak et al.(1933)) in Riemannian  space ~$R_n$~ of dimensions (n), whose
    metric is given by:
     \begin{equation}
    dS^2=g_{\mu\nu}~dx^\mu~dx^\nu.
    \end{equation}
   Consider the null-geodesic $~\Gamma~$ connecting the two neighboring $ C_0 $ and $ C_1 $ in $ R _n $. The tangent null-vector
   ( transport  vector in (Kermak et al.(1933))) is defined by,

   \begin{equation}
   \eta^{\rho} \edf \frac{dx^\rho}{d\lambda},
   \end{equation}
   where $~\lambda~$ is a parameter characterizing $~\Gamma~$. The equation for
   $~\Gamma~$ is given by,
    \begin{equation}
   {\frac{d\eta^\rho}{d~\lambda}}
   +\cs{\mu}{\nu}{\rho}~{\eta^\mu}{\eta^\nu}=0,
   \end{equation}
   where $\cs{\mu}{\nu}{\rho}$ is the Christoffel symbol of the second type. Equation (3) follows from the Euler-Lagrange equation,
   \begin{equation}
   \frac{d}{d\lambda}(\frac{\partial~T}{\partial~\eta^\mu})
    -\frac{\partial~T}{\partial~x^\mu}=0,
   \end{equation}
   upon taking,
    \begin{equation}
    ~T ~\edf~\frac{1}{2}~g_{\mu\nu}~\eta^\mu~\eta^\nu~.
    \end{equation}

    Let $~ \Gamma^{'}~ $ be a null-geodesic, parallel to $~\Gamma~$ and
    passing through the point$ ~C^{'}~$ near C, and let $~\xi^\sigma~$
    denotes the vector$~ CC{'}~$. Let a scalar J  be defined as,

   \begin{equation}
   J \edf \eta_{\alpha}~{\xi^{\alpha}} ,
   \end{equation}
   where $~\eta_\alpha~$is the covariant form of the vector $~\eta^\alpha~$.

 The  KMW -Theorems can be stated as,\\

{\subsection*{\bf {Theorem (I)}}}

{\it{"The scalar J given by (6) is independent of
 the choice of the direction  CC', and is also independent of the
 position of C on the null-geodesic $\Gamma$. "}}\\
 It depends only on the two null-geodesics  as a whole and not on any
 particular point on them. This is the first theorem which was
 rigorously proved by KMW.

  Consider a null-geodesic parallel to $\Gamma$, of which there
  are $~\infty ^{n-1}~$. If we form the scalar "J" corresponding
  to $\eta^\mu $ and any of these null-geodesic, we find that for
  any particular value of "J" there are$~\infty ^{n-2}~$ parallel
  null-geodesics.\\

{\subsection*{\bf  {Theorem (II)}}}

{\it {"If the set of $~\infty ^{n-2}~$ null-geodesics lies in a
$~R_{n-1}~$
   which intersects a local  flat subspace $~E_{n-1}~$,
   at C, in a local flat subspace $~E_{n-2}~$ , then
   $~E_{n-2}~$ is perpendicular to the projection of $~\Gamma~$ in
   $~E_{n-1}~$."}}

   As a consequence of these two theorems KMW were able to derive
   the following formula for the red-shift extracted from
   photons, assuming that a photon is moving along null-geodesic of the metric,
   \begin{equation}
Z={\frac{\Delta ~\lambda}{\lambda}}=
{\frac{[{\rho}_{\mu}~{\eta}^{\mu}]_{C_1}~-~{[{\varpi}_{\mu}~{\eta}^{\mu}]_{C_0}}}
{[{\varpi}_{\mu}~{\eta}^{\mu}]_{C_0}}}.
\end{equation}
   where  $~\rho_\mu, \varpi_\mu~$
are the covariant form of the tangents to the geodesics of the
observers at  $~C_1, C_0~$ respectively.
\section{Spin-Dependent Path Equation}
It is well known that  Riemannian geometry possesses two types of
the paths.
   The first is the geodesic path and the second is the null-geodesic.
     The equation of these two paths can be written in the general form,
   \begin{equation}
   {{\frac{d^2x^\mu}{d~p^2}} +
   \cs{\alpha}{\beta}{\mu}~{\frac{dx^\alpha}{d~p}}~{\frac{dx^\beta}{d~p}}=
   0},
   \end{equation}
   where $p$ is a  parameter characterizing the trajectory of massive
     or massless particle (cf. Adler et al.(1975)). This parameter may be related
     to the parameter $S$, of (1), by,

   \begin{equation}
   dS^2 = E dp^2
   \end{equation}
   where $E$ is a numerical parameter taking the values
   \begin{eqnarray}
   E&=&0,~for ~a~ null-geodesic ~,\nonumber\\
   E&=&1, ~~~for~a ~geodesic~.
   \end{eqnarray}

   Wanas et al. (1995), directed their attention to the absolute parallelism space
   (AP-space), and by generalizing the method given by Bazanski (1977), (1989),
   they derived the following set of three path equations :
   \begin{equation}
   {{\frac{dJ^\mu}{dS^-}} + \cs{\nu}{\sigma}{\mu}\ J^\nu J^\sigma = 0},
   \end{equation}
   \begin{equation}
   {{\frac{dW^\mu}{dS^o}} + \cs{\nu}{\sigma}{\mu}\ W^\nu W^\sigma =-
   {\frac{1}{2}}\Lambda_{(\nu \sigma)} . ^\mu~~ W^\nu W^\sigma},
   \end{equation}
   \begin{equation}
   {{\frac{dV^\mu}{dS^+}} + \cs{\nu}{\sigma}{\mu}\ V^\nu V^\sigma =-
   \Lambda_{(\nu \sigma)} . ^\mu~~ V^\nu  V^\sigma} ,
   \end{equation}
   where $J^\mu$ , $W^\mu$ and $V^\mu$ are the tangents to the corresponding
   curves whose parameters are $S^-$, $S^0$ and $S^+$ respectively,
   and $\Lambda^\alpha _{~\mu \nu}$ is the torsion of the AP-geometry defined by,
   \begin{equation}
   \Lambda^\alpha_{~ \mu \nu} \edf ~ \Gamma^\alpha_{. \mu \nu}
   -\Gamma^\alpha_{. \nu \mu},
   \end{equation}
   where $\Gamma^\alpha_{. \nu \mu}$ is the non-symmetric affine connection
   defined as a consequence of the condition for AP.

   Wanas (1998) defined a general expression for a connection formed by    taking linear
   combinations of the available connections in the AP-space. He mentioned that the
   metricity condition is necessary but not sufficient to define the
   Christoffel symbol. He generalized the three path equations (11), (12) and (13) in the following equation,
   \begin{equation}
   {{\frac{dZ^\mu}{d\tau}} + \cs{\nu}{\sigma}{\mu}\ Z^\nu Z^\sigma =-
   {b}\Lambda_{(\nu \sigma)} . ^\mu~~ Z^\nu Z^\sigma},
   \end{equation}
   where b is a parameter given by $b={\frac{n}{2}}{\alpha}{\gamma}$,
$n(=0,1,2,...,)$ is a natural number $\alpha~$ is the fine structure
constant and $\gamma~$
 is a numerical free parameter,
   $Z^{\mu}{\edf}{\frac{d~X^\mu}{d~\tau}}$,
   and $\tau $  is the evolution
   parameter along the new general path (15) associated with the general connection:
   \begin{equation}
   \nabla^\alpha_{.~\mu \nu }=
   ~\cs{\mu}{\nu}{\alpha}+~b~ \gamma^\alpha_{. ~\mu
   \nu} ,
   \end{equation}
The geometric structure characterized by the connection (16) is
called the PAP geometry (Wanas(2000)). It is worth of mention that
equation (15), represents a
   generalization of the three path equations given above. This
   equation will reduce to the equation of geodesic (null-geodesic upon
   reparameterization) in the Riemannian geometry, when the parameter $b=0$.

  It has been shown that (15), can be used to represent trajectories of spinning test particles, massive or
  massless, in a gravitational field.

\section{Theorems on Null-Paths in PAP-spaces}

Let $~{\Gamma}~$, $~{\Gamma}^{'}~$ be two neighboring null-paths of
the type given by (15) defined in PAP-space of dimensions $(n)$. Let
C be a point on $~{\Gamma}~$ and $~{C^{'}}~$ be a neighboring point
on $~{\Gamma~}'~$. Let $~\zeta^\mu~$ be the vector $~{CC^{'}}~$
connecting the two points. Let us define the following scalar,
  \begin{equation}
   ~{J}{\edf} Z^{\mu}{\zeta_\mu},
   \end{equation}
where $~Z^\mu~$ is the null tangent to the path
(15), defined at C. Then we can prove the following theorem:\\

   {\subsection*{\bf {\it {Theorem (I)}}}}
   {\it {"The scalar $~{J}$ is independent of the position of
    the point C on the null-path $~\Gamma~$ and is also
    independent of the choice of the direction $~CC^{'}~$ but depends on the two null-paths themselves."}}\\
Now, as  mentioned in the previous section, the equation of the null
paths in the PAP Geometry is given by (15) and the equation of
 null geodesic in The Riemannian Geometry is
   given by (8). It is clear  that equation (15) tends to equation (8), if$~b=0~$.
    Keeping in mind the fact that for every absolute parallelism
    space there  exists an associated Riemannian one, then we can
    relate the objects in the first equation (15) to the objects of the
    second (8) by the following relations

   \begin{equation}
   Z^\nu=\eta^\nu~(1+g(b)),
   \end{equation}
\begin{equation}
{\tau}=p~(1+f(b)),~and
\end{equation}
\begin{equation}
\zeta_\nu=\xi_\nu~(1+l(b)).
   \end{equation}
where $g(b), f(b)$ and $l(b)$ are  positive functions of the
parameter b such that these functions tend to zero when $b$ goes to
zero. Now let us evaluate the scaler$ ~{J}{\edf}
Z^{\mu}{\zeta_\mu}~$. By using equation (19), we get
\begin{equation}
\frac{d~p}{d~{\tau}}=\frac{1}{(1+f(b))}
\end{equation}
If we use (18),(20) and (21), we have
\begin{eqnarray}
\frac{d~}{d~{\tau}}(Z^{\mu}{\zeta_\mu})&=& \frac{1}{(1+f(b))}{\frac{d~}{d~p}}(Z^{\mu}{\zeta_\mu})\nonumber\\
 &=& \frac{1}{(1+f(b))}({\frac{d~}{d~p}}(\eta^\nu~(1+g(b))(\xi_\nu~(1+l(b))))\nonumber\\
&=&
\frac{1}{(1+f(b))}(1+g(b))(1+l(b))({\frac{d~}{d~p}}(\eta^\nu~\xi_\nu~))
\end{eqnarray}
Now by recalling equation (6) and theorem(I) of KMW, we get
\begin{eqnarray}
\frac{d~}{d~{\tau}}(Z^{\mu}{\zeta_\mu})&=& 0, \nonumber\\
i.e~~~~{Z^\mu}{\zeta_\mu}&=&constant.
\end{eqnarray}

      This result proves theorem (I) on the null-path of the PAP-space. Now,
   as an extension of the idea of null-path$~\Gamma~$ passing through a given
   point C in the PAP-space $~T_n~$, one can find $~\infty ^{n-1}~$ of
  null-paths, in the neighborhood of the point C, parallel to the
   null-path$~\Gamma~$.  The second theorem can be stated as
   follows:\\
   {\subsection*{\bf {\it {Theorem (II)}}}}
   {\it {For a definite value of the scalar J, we have $~\infty ^{n-2}~$
   of parallel null-paths lying in a subspace $~T_{(n-1)}~$, which
   intersects
   a local flat subspace $~E_{(n-1)}~$ of (n-1) dimensions at the point C
    in a local (n-2)-dimensions flat subspace$~E_{(n-2)}~$. This $~E_{(n-2)}~$
     is perpendicular to the projection of the null-path $~\Gamma~$
   in $~E_{(n-1)}~$.}}\\

   To prove this theorem, we shall assume that at the point C, or at
   any point in its neighborhood, there is no singularity. If we assume
   any rectangular axes at C, such that
   the direction ratios along the null-path $~\Gamma~$ are
   $~(\lambda_1,\lambda_2,......,\lambda_n)~$, then:
   \begin{equation}
   {\lambda_1}^2+{\lambda_2}^2+......+{\lambda_n}^2=0.
   \end{equation}
   So, the tangent null-vector $Z^\mu~$, to the
   null-path at C, has the components, $~(m\lambda_1,m\lambda_2,......,m\lambda_n)~$,
  where $(m)$ is some
   constant. If the coordinates of the point C are $~(x^1,x^2,....,x^n)~$, and if we
   take the vector $~Z^\mu~$, which is defined above, as
   $~(x^1,x^2,...,x^n)~$ , then the scalar J takes the form:
   \begin{equation}
   J=(x^1\lambda_1+x^2\lambda_2+...+x^n \lambda_n)m,
   \end{equation}
  i.e
   \begin{equation}
   (x^1\lambda_1+x^2\lambda_2+....+x^n\lambda_n)=\frac{~{J}}{m}.
   \end{equation}
    Due to theorem (I), J remains constant wherever $~(x^1,x^2,...,x^n)~$
   lies in the hyperplane, i.e, for any other point$~C^{'}~$on null
   path $~\Gamma^{'}~$, parallel to$~\Gamma~$, of coordinates
    $~(x^1+k~\lambda_1,x^2+k~\lambda_2,...,x^n+k~\lambda_n)~$, where
   k is a variable parameter depending on the null-path
$~{\Gamma^{'}}~$, J
    remains constant. So, if we take
   the components of $~Z^\mu~$ at point $~C^{'}~$ on the null-path
   $~\Gamma^{'}~$,to be
   $~(x_1+k\lambda_1,x_2+k\lambda_2,....,x_n+k\lambda_n)~$,and
   substitute in equation (28), we have:\\
   $$~{J}=m(x^1\lambda_1+k{\lambda_1}^2+x^2\lambda_2+k{\lambda_2}^2+.....+x^n\lambda_n+k{\lambda_n}^2).$$
   Then,  using (24), we get the same equation (25). This means that  any
   one of the set of the  null-paths parallel to the null-path$~ \Gamma~$,
   must lie in the hyperplane  given by (26), in order to keep $ ~{J} $ constant. Now if we put $~x^1=0~$, it follows
   directly
   that all the points in which these null-paths cut any local subspace
   $~E_{(n-1)}~$ lie in a local subspace $~E_{(n-2)}~$  given by:
   $$ (x^2\lambda_2+.....+x^n\lambda_n)=\frac{~{J}}{m}.$$
   This $~E_{(n-2)}~$ is perpendicular to the null-path $~\Gamma~$ whose
   projection in $~E_{(n-1)}~$ is given
   by:

   \begin{equation}
   \frac{x^2}{\lambda_2}=.....=\frac{x^n}{\lambda_n}.
   \end{equation}
Hence, the second theorem is proved.

      By using the two previous theorems and the same idea of KMW, we can
   write a general expression of the red-shift. This expression depends
   essentially on the idea of the wave fronts,  which are represented
   by a set of null-paths passing through the points of the wave front.
Their projections in the local subspace are perpendicular to this
   wave front. So, the actual wavelength is determined by the perpendicular
   distance between the wave fronts corresponding to two parallel sets of
   null-paths. In other words, it is the interval between the
   points of intersection of the two local subspaces defined by the two sets of parallel null-paths and the observer's world line.

\section{General Expression of Red-Shift in PAP-Space}
In the previous section, we have shown that the two theorems on
   null-geodesics, proved by KMW, are also applicable to the
   null-paths (15). Now , we are going to assume that the trajectory of a photon
   (spin-1 particle) in a gravitational field is spin dependent and given by equation
   (15).
   So, we can easily establish a general formula for the red-shift
   of spectral lines similar to that given by KMW.

      Consider a null path of the form (15) connecting the two
      points $~C_1~$, $~C_0~$ at which two observers A and B are
      located, respectively. The null paths $~\Gamma~$ belongs to
      the same wave front observed by A and B. Let
      $~{\eta_1}^{\mu}~$ and $~{\eta_0}^{\mu}~$ be the components
      of the transport null tangent to$~\Gamma~$at$~C_1~$ and
      $~C_0~$, respectively. Let$~\Gamma^{'}~$ be a null path
      parallel to $~\Gamma~$ , belonging to the succeeding  wave front,
      intersecting the world lines of A and B at$~C_1'~$ and
      $~C_0'~$,
      respectively. If $\lambda_1~$, $\lambda_0~$ are the
      wavelengths of the same spectral line as observed by A and B,
      respectively, then the components of the
      vectors$~C_1{C_1}'~$and$~C_0{C_0}'~$ are
      $~\rho^{\mu}~,\varpi^{\mu}~$,which represent the components of the
      unit tangents to the world lines of A and B, respectively.
      These unit tangents are solutions of (15), for A and B, upon
      taking $b=0$. The vectors $\rho^{\mu}$ and $\varpi^{\mu}$ are the
      values of the vector
      $~\xi^{\mu}$, of theorem (I), evaluated at$~C_1~$, $~C_0~$
      respectively, while $~{\eta_1}^{\mu}~$, $~{\eta_0}^{\mu}~$
      represent the
      values of the vector $~Z^\mu~$, of the same theorem, evaluated
      at$~C_1~$, $~C_0~$as stated above. Now applying theorem (I)
      and equating the values of $~{J}$ at$~C_1~$, and at  $~C_0~$we get

      \begin{equation}
   \lambda_1~{\eta_1}^{\mu}~\rho_\mu=\lambda_0~{\eta_0}^{\mu}~\varpi_\mu
   \end{equation}
\begin{equation}
 i.e.  \frac{\lambda_0}{\lambda_1}
   =\frac{{\rho}_{\mu}~{{\eta}^{\mu}}_1}{{\varpi}_{\mu}~{{\eta}^{\mu}}_0},
   \end{equation}
thus,
\begin{equation}
   \frac{\Delta\lambda}{\lambda_1}= \frac{\lambda_0-\lambda_1}{\lambda_1}
   ={\frac{{\rho}_{\mu}~{{\eta}^{\mu}}_1}{{\varpi}_{\mu}~{{\eta}^{\mu}}_0
   }}-1.
   \end{equation}
   This gives a general formula for the red-shift of spectral
   lines coming from a distant object.
\section{Concluding Remarks}
In the present work,we have investigated the validity of two
important theorems on null-geodesics in the context of the PAP
-geometry. These two theorems are important to establish a general
formula for the red-shift of spectral lines especially when the
trajectories of massless particles are spin dependent. Equation (15)
is the equation representing such trajectories .This equation can be
used as an equation of motion in the context of any field theory
written in the AP-geometry including GR (cf. Wanas (1990)).

In conclusion we can write the following general remarks\\
 (1) In the present work, we tried, as
far as we could, to use the same notations, as those used in the
original work of KMW in order to facilitate comparison.\\ (2) The
path equation (15) can be used to represent the trajectory of a test
particle or the trajectory of a massless particle, in a background
gravitational field, upon adjusting the parameter $~\tau~$. The
right hand side of this equation is suggested to represent a type of
interaction between the spin of the moving particle and the torsion
of the background gravitational field. The
parameter $(b)$ is a spin dependent parameter (Wanas (1998)).\\
(3) The two theorems on (15), when it represents  null paths, can be
reduced to the original KMW theorems, reviewed in section 2, upon
taking $b=0$.\\ (4) Equation (30) gives the red-shift taking into
account the spin-torsion interaction. This equation appears to be
the same as that given by KMW, but the main difference is that the
effect of the spin of the moving particle will appear in the values
of the null-vectors$~{\eta_0}^{\mu}~$and $~{\eta_1}^{\mu}~$ which
are
solutions of equation(15) and not of the equation of null-geodesic.\\
 (5) In
KMW-paper, the authors used formula (7) to get Doppler shift. In an
attempt to interpret the solar limb effect (Mikhail, et al. (2001)),
the use of (7) have been widened to account, not only for the
relative radial
velocity between A and B (Doppler-shift), but also for:\\
 (i) The effect of gravity (gravitational red-shift).\\
 (ii) The effect of the direction of the null-geodesic.\\
 In addition to these effects, the formula (30) will account also
 for the effect of the spin-torsion interaction on the value of the
 red-shift.\\
 (6) Formula (30) can be used to get the red-shift whether it
 is treated as a metric phenomena or not.

\end{document}